\documentclass[dvipsnames]{llncs}
\usepackage{todonotes}
\usepackage{hyperref}
\usepackage{booktabs}
\usepackage[figuresright]{rotating}
\usepackage{listings}
\usepackage[T1]{fontenc}                              
\usepackage[scaled=0.80]{beramono}
\usepackage{lscape}
\usepackage{longtable}
\lstdefinelanguage{Pseudocode}{
  morekeywords=[1]{for, each, print, do, done},
  morekeywords=[2]{Type, Prop, Set, true, false, option},
  morecomment=[s]{(*}{*)},
  sensitive=true,
  commentstyle=\color{gray},
  identifierstyle={\ttfamily\color{black}},
  keywordstyle=[1]{\ttfamily\color{violet}},
  keywordstyle=[2]{\ttfamily\color{blue}}
}
\usepackage{xspace}
\newcommand{\nTotalBenchmarks}{105\xspace}
\newcommand{\nDaisyAlarms}{34\xspace}
\newcommand{\nDaisyGood}{48\xspace}
\newcommand{\nDaisyAlarmsInput}{24\xspace}
\newcommand{\nDaisyAlarmsOutput}{27\xspace}
\newcommand{\nHerbieTimeouts}{24\xspace}
\newcommand{\nTimeoutMinutes}{2\xspace}
\newcommand{\nHerbieResults}{48\xspace}
\newcommand{\nHerbieBetter}{23\xspace}
\newcommand{\nHerbieEqual}{14\xspace}
\newcommand{\nHerbieWorse}{11\xspace}
\newcommand{\nHerbieNotWorse}{37\xspace}
\newcommand{\nDaisyBetter}{33\xspace}

\newcommand{\nDaisyWorse}{1\xspace}

\newcommand{\nBothBetter}{30\xspace}

\newcommand{\nSomeSucceed}{71\xspace}
\newcommand{\daisyRWFactor}{1.2 (0.08 orders of magnitude)\xspace}
\newcommand{\herbieRWFactor}{10.14 (1.01 orders of magnitude)\xspace}
\newcommand{\bothRWFactor}{12.79 (1.11 orders of magnitude)\xspace}

\title{Combining Tools for Optimization and Analysis of Floating-Point Computations}
\author{Heiko Becker\inst{1} \and Pavel Panchekha \inst{2} \and Eva Darulova \inst{1}
   \and Zachary Tatlock \inst{2}}
\institute{
  MPI-SWS \{\href{mailto:hbecker@mpi-sws.org}{hbecker},\href{mailto:eva@mpi-sws.org}{eva}\}@mpi-sws.org \and
  University of Washington \{\href{mailto:pavpan@cs.washington.edu}{pavpan},\href{mailto:ztatlock@cs.washington.edu}{ztatlock}\}@cs.washington.edu}

\begin{document}
\maketitle

  \begin{abstract}
    Recent renewed interest in optimizing and analyzing floating-point
    programs has lead to a diverse array of new tools for numerical programs.
    These tools are often complementary, each focusing on a distinct aspect of
    numerical programming.  Building reliable floating point applications
    typically requires addressing several of these aspects, which makes easy
    composition essential. This paper describes the composition of two recent
    floating-point tools: Herbie, which performs accuracy optimization, and
    Daisy, which performs accuracy verification. We find that the combination
    provides numerous benefits to users, such as being able to use Daisy to check
    whether Herbie's unsound optimizations improved the worst-case roundoff
    error, as well as benefits to tool authors, including uncovering a number of
    bugs in both tools. The combination also allowed us to compare the different
    program rewriting techniques implemented by these tools for the first time.
    The paper lays out a road map for combining other floating-point tools and
    for surmounting common challenges.

  \end{abstract}
  %
\section{Introduction}\label{sec:introduction}

  Across many domains, numerical computations specified over the reals are actually
  implemented using floating-point arithmetic. Due to their finite nature,
  operations on floating-point numbers cannot be calculated exactly
  and accumulate
  roundoff errors. In addition, real-valued identities such as associativity no
  longer hold, making manual reasoning and optimization challenging. To address
  these challenges, new automated tools have recently been developed which
  build on advances in program rewriting and verification techniques to enable
  even non-experts to analyze and optimize their floating point code.


  \begin{sloppypar}
  Some of these tools use sound techniques to statically bound roundoff errors
  of straight-line floating-point
  programs~\cite{Rosa,Fluctuat,FPTaylor,real2Float,Precisa,Daisy} and partially
  automate complex analysis tasks~\cite{VCFloat,Gappa}. Other such tools use
  dynamic techniques to find inputs that suffer from large rounding
  errors~\cite{chiang2014efficient,zou2015genetic}. Yet other tools perform
  rewriting-based optimization~\cite{salsa,Panchekha2015,sanchez2017,Daisy} and
  mixed-precision tuning~\cite{FPTuner,Darulova2018} to improve the accuracy and
  performance of floating-point programs.
  \end{sloppypar}

  Since these tools are typically complementary, each focusing on a distinct
  aspect of numerical reliability, users will need to compose several to
  meet their development needs. This makes ease of composition essential, and
  some first steps in this regard have been taken by the FPBench
  project~\cite{FPBench}, which provides a common specification language for
  inputs to floating-point analysis tools similar to the one provided by the
  SMT-LIB standard~\cite{SMTLIB}. However, no literature yet exists on the
  actual use of FPBench to compose tools and on the
  challenges that stand in the way of combining different floating-point tools,
  such as differing notions of error and different sets of supported functions.

  In this paper we report on our experience implementing the first
  combination of two complementary floating-point analysis tools using FPBench:
  Herbie~\cite{Panchekha2015} and Daisy~\cite{Daisy}.
  Herbie optimizes the accuracy of straight-line floating-point expressions, but
  employs a dynamic roundoff error analysis and thus cannot provide sound
  guarantees on the results.
  In contrast, Daisy performs static analysis of straight-line expressions,
  which is sound w.r.t. IEEE754 floating-point semantics~\cite{IEEE754}.
  Our combination of the tools is implemented as a script in the FPBench repository
  (\url{https://github.com/FPBench/FPBench}).

  We see this combination of a heuristic and a sound technique as
  particularly interesting; Daisy can act as a backend for validating Herbie's
  optimizations. Daisy computes improved worst-case roundoff error bounds for
  many (but not all) expressions optimized by Herbie. On others it raises an
  alarm, discovering division-by-zero errors introduced by Herbie. We also
  improved the precision of Daisy's analysis of special functions as we found
  that some were sound but not accurate enough. Thus, the combination was also
  useful in uncovering limitations of both tools.

  Daisy additionally implements a sound genetic programming-based accuracy
  optimization procedure. Our combination of Daisy and Herbie allows us to
  compare it to Herbie's unsound procedure based on greedy search. We discover
  important differences between the two procedures, suggesting that the
  techniques are not competitive but in fact complementary and best used in
  combination.

  Some of the challenges we encountered, such as differing supported functions
  and different error measures, are likely to be encountered by other
  researchers or even end users combining floating-point tools, and our
  experience shows how these challenges can be surmounted. Our evaluation on
  benchmarks from the FPBench suite also shows that tool composition can provide
  end-to-end results not achievable by either tool in isolation and suggests
  that further connections with other tools should be investigated.


  %
\section{Implementation}\label{sec:implementation}

The high-level goal of our combination is to use Daisy as a verification backend
to Herbie to obtain a sound upper bound on the roundoff error of the expression
returned by Herbie.
By also evaluating the roundoff error of Herbie's output and of the input
expression, we can obtain additional validation of the improvement. It should be
noted, however, that Daisy cannot verify whether the actual worst-case or
average roundoff error has decreased---a decrease in the computed upper bound
can be due to an actual decrease or simply due to a stronger static
bound. In many cases, however, such as in safety-critical systems,
just \emph{proving} a smaller static bound is already useful.

We have implemented the combination in a
script, which we sketch in
\autoref{fig:Daisy-Herbie-flow}. For each straight-line input program
\lstinline{f}$_{src}$, we first run Herbie to compute an optimized version
\lstinline{f}$_{res}$. Both the optimized and unoptimized version are translated
into Daisy's input format (using \lstinline{FPCore2Scala}), and Daisy is run on
both versions to compute error bounds.

Daisy supports several different types of error analysis, and we run Daisy in a
portfolio style, where the tightest bound computed by any of the analyses is
used. In particular, we use the interval analysis with subdivisions mode and the
SMT solver mode (with Z3~\cite{Z3} as the solver).\footnote{We found that
  neither interval analysis without subdivision nor alternate SMT solvers
  provided tighter bounds.} Since each analysis is sound, this provides the
tightest error bound that Daisy can prove.

To implement the toolchain from \autoref{fig:Daisy-Herbie-flow}, we had to
address two major differences between Herbie and Daisy: Herbie and Daisy use
different input (and output) formats, and Daisy requires domain bounds on all input
variables, whereas Herbie allows unbounded inputs.

\begin{figure}[t]
  \begin{lstlisting}[language=Pseudocode, escapechar=|]
f|$_{res}$| = Herbie(f|$_{src}$|)
err|$_{src}$| = min{ Daisy(A, FPCore2Scala(f|$_{src}$|)) |$\mid$| A <- AnalysisTypes }
err|$_{res}$| = min{ Daisy(A, FPCore2Scala(f|$_{res}$|)) |$\mid$| A <- AnalysisTypes }
  \end{lstlisting}
  \caption{Pseudocode of the script used to compose Herbie and Daisy into a
    single tool; \lstinline{AnalysisTypes} contains different modes Daisy can be
    run in.}
  \label{fig:Daisy-Herbie-flow}
\end{figure}

\paragraph{Formats}
To avoid having to add new frontends, we implemented a translator from
FPBench's FPCore format to Daisy's Scala-based input language. As Herbie
produces optimized expressions in FPCore, this translator allows us to run Daisy
on both the benchmarks and on Herbie's optimized expressions.
This translator is now part of the FPBench toolchain and can be used by other
researchers and by users to integrate Daisy with other tools developed as part
of the FPBench project.

\paragraph{Preconditions}
Both Daisy and Herbie allow preconditions for restricting the valid
inputs to a floating-point computation. For Herbie, these preconditions are
optional. In contrast, Daisy requires input ranges for performing a forward
dataflow analysis to compute sound absolute roundoff error bounds.
Several of the benchmarks in FPBench did not have a specified precondition. For
our experiments, we manually added a meaningful precondition to these programs,
with preconditions chosen to focus on input values with significant rounding
errors. To avoid biasing the results, the preconditions were simple order-of
magnitude ranges for each variable, with the endpoints of these ranges chosen
from $1$, $10^{10}$, or $10^{20}$ and their inverses and negations.

\paragraph{Improvements in Daisy and Herbie}

Connecting Daisy and Herbie and running each on several previously unseen
benchmarks uncovered numerous possibilities for improvements in Herbie and
Daisy.

In Herbie, several bugs were discovered by our efforts: an incorrect
type-checking rule for \textsf{let} statements (which would reject some valid
programs); incorrect handling of duplicate fields (which allowed one field to
improperly override another); and an infinite loop in the sampling code
(triggered by some preconditions). Real users running older versions of Herbie
have since also reported these bugs, suggesting that issues addressed during
tool composition helped improve user experience generally.

In Daisy, we discovered that the analysis of elementary functions was
unnecessarily conservative and improved the rational approximations used. Error
handling in both tools was also improved such that issues like (potential)
divisions by zero or timeouts are now accurately reported. This more precise
feedback significantly improves user friendliness and reduces debugging time.
\section{Experimental Results}\label{sec:results}

We perform two evaluations of our combination of Daisy and Herbie: we first
use Daisy as a verification backend for Herbie and then
we compare Daisy's and Herbie's rewriting algorithms.
Both experiments use all supported benchmarks from the
FPBench suite. A table with all the evaluation data can be found in the appendix.
The experiments were run on a machine with an i7-4790K CPU and 32GB of
memory. For each benchmark we give both Daisy and Herbie a timeout of
\nTimeoutMinutes minutes.

\begin{figure}[t!]
  \input{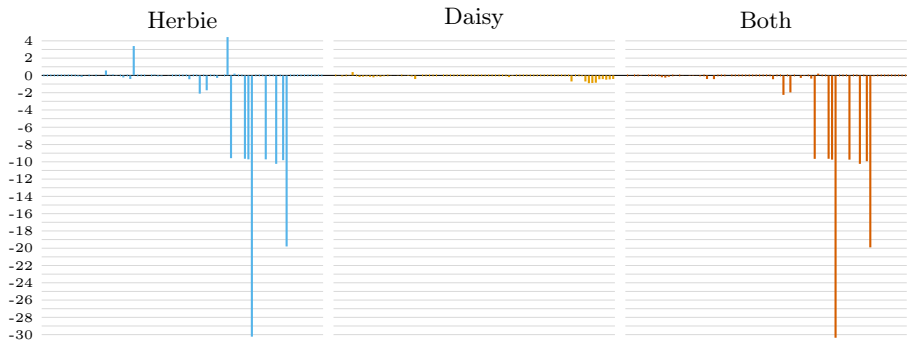}
  \caption{The orders of magnitude change in Daisy's worst case error
    estimate after rewriting with Herbie, Daisy, and both Herbie and Daisy
    (left to right). Note that combining both
    rewriting algorithms keeps the large beneficial changes introduced
    by Herbie but avoids its detrimental changes.}
  \label{fig:results}
\end{figure}

\paragraph{Composing Daisy and Herbie}
Our first experiment considers Daisy solely as a tool for computing
floating-point error bounds. Herbie is then used to attempt to improve
the benchmark's accuracy.

Of the \nTotalBenchmarks benchmarks, Herbie times out on \nHerbieTimeouts of them. Of
the remaining \nHerbieResults benchmarks, Daisy raises an alarm%
\footnote{Indicating that it could not prove the absence of invalid
  operations, such as divisions by zero.}
on \nDaisyAlarms and can prove a
bound on \nDaisyGood. Of the \nDaisyGood benchmarks where Daisy can prove an error
bound, Daisy's roundoff error analysis can prove a tighter worst-case
error bound for \nHerbieBetter of Herbie's outputs, an equal bound for \nHerbieEqual, and a
looser bound for \nHerbieWorse.
These results are summarized in the left-most graph in~\autoref{fig:results}.

Of the \nDaisyAlarms benchmarks where Daisy raises an alarm, for \nDaisyAlarmsInput Daisy raises
an alarm on the original input program and for \nDaisyAlarmsOutput the alarm is raised
on the output program Herbie produced. Some of these alarms are true positives while
others are spurious. For example, in some benchmarks Herbie had
introduced a possible division by 0, which Daisy was able to detect.
In others, Herbie's output contained expressions like $a / (1 + b^2)$,
where Daisy is unable to prove that no division by zero occurs.

We see this as good evidence that Daisy can be used as a verification
backend for Herbie. One challenge is that Daisy's error
analysis can only show that Herbie's output has a smaller error \emph{bound},
not that it is more accurate at any particular point on the
original program. Additionally, it is difficult to determine which of
Daisy's alarms are spurious. Despite these challenges, the combination
of Daisy and Herbie was able to produce large, verified improvements
in rounding error in many benchmarks.

\paragraph{Comparing Daisy's and Herbie's Error Measures}
One topic of particular interest in combining Daisy and Herbie is
their different measures of floating-point error. Herbie measures the
error in terms \textit{units in the last place}, or ULPs. To compute
this error, Herbie randomly samples input values and takes the average
of error across those inputs. It thus provides a \textit{dynamic,
  unsound measure of average ULPs} of error. Daisy, meanwhile, uses a
mathematical abstraction based on the definition of rounding and
IEEE754 operator semantics to provide \textit{static, sound bounds on
  the maximum absolute error}. The relationship between these two
measures of error is central to using Herbie and Daisy together.

Despite these stark differences between Herbie's and Daisy's error
measures, our evaluation data shows that Daisy and Herbie can be
fruitfully used together: Daisy verifies that Herbie's improved
program is no less accurate for \nHerbieNotWorse/\nHerbieResults of
the benchmarks. This suggests that, though Daisy and Herbie use very
different means to measure error, both are successfully measuring the
same underlying notion of error. The fact that Daisy's and Herbie's
error measures are suited to their particular approaches (static
analysis and program search) suggests that future tools should focus
not on measuring error ``correctly'' but on finding an error measure
well suited to their technical approach.

\paragraph{Comparing Daisy and Herbie}\label{sec:results-rewriting}

Our second experiment compares Daisy's and Herbie's rewriting algorithms. Daisy
uses genetic programming to search for a program with a tighter bound. Herbie,
by contrast, uses a greedy search over a suite of different rewriting steps.
We compare Herbie's rewriting algorithm, Daisy's rewriting algorithm,
and Daisy's rewriting algorithm applied to the results of Herbie's
rewriting algorithm. \autoref{fig:results} summarizes the accuracy improvements.

Of the \nTotalBenchmarks benchmarks, at least one of the rewriting algorithms succeeds on \nSomeSucceed.
Of the \nSomeSucceed, Daisy's rewriting algorithm tightens the worst-case bound
for \nDaisyBetter benchmarks; Herbie's for \nHerbieBetter benchmarks; and the combination for
\nBothBetter benchmarks. Furthermore, Herbie's rewriting algorithm loosens the
worst-case bound for \nHerbieWorse benchmarks, a consequence of its unsound error
measurement technique or differing notion of error, while the
combination does so for only \nDaisyWorse.

Not only the number but also size of the error improvement matters.
Daisy's rewriting algorithm was able to reduce the error bound by a
factor of \daisyRWFactor on average; Herbie's by a factor of
\herbieRWFactor; and the combination by a factor of \bothRWFactor.
The combination clearly provided the greatest reduction in error bounds;
furthermore, Daisy's algorithm provides larger benefits when applied to Herbie's
optimized program than when applied to the benchmark directly.

It seems that Daisy's rewriting algorithm provides a fairly consistent
but small tightening of error bounds, while Herbie's algorithm can
suggest dramatic and unexpected changes in the expression. However,
these large changes sometimes have significantly \textit{looser} error
bounds. In those cases, combining Herbie's and Daisy's rewriting
algorithms provides a tighter error bound, reaping the benefits of
Herbie's rewriting algorithm without the large increases in error
bounds that it sometimes causes.


  %
  \section{Discussion}

This paper reports on the combination of Daisy and Herbie and
illustrates the benefits of composing complementary floating-point
tools to achieve results neither tool provides in isolation. This case
study serves as a representative example: similar combinations could
be constructed for other tools using this paper's approach.
Combinations of Gappa~\cite{Gappa}, Fluctuat~\cite{Fluctuat},
FPTaylor~\cite{FPTaylor}, or other verification
tools~\cite{Rosa,real2Float,Precisa} with Herbie could also allow
validating Herbie's optimizations. Verification tools could also be
used to validate the output of other unsound tools, such as
Precimonious~\cite{Precimonious} and STOKE~\cite{Stoke}. Comparisons
with sound optimization tools such as Salsa~\cite{salsa} and
FPTuner~\cite{FPTuner} could also be explored.

In our experiments we have seen that there is no correlation between
the error measures of Daisy and Herbie. Experiments with other tools
and other error measurements could help clarify the relationships
between different error measures. Particularly interesting is whether
other uncorrelated error measures nonetheless evaluate changes to
programs in correlated ways, as Daisy's and Herbie's error measures do.
Tools that use error measures related in this way can be composed,
even if they use different error measures

Ultimately, we envision using the combination of Daisy and Herbie within
larger developments such as VCFloat~\cite{VCFloat}. VCFloat provides
partial automation for reasoning about floating-point computations in
CompCert C-light~\cite{Leroy-Compcert-CACM} programs. In this context,
our toolchain could provide an optimization tactic, that
could be applied to (provably) increase accuracy for floating-point
segments of C-light programs.


\bibliographystyle{splncs03}
\bibliography{bibliography}

\appendix
\newpage
\section{Experimental Data}\label{sec:experiment_data}

In this section we give the experimental results on which all conclusions from
\autoref{sec:results} are based. The experiments were run on a machine with an
i7-4790K CPU and 32GB of memory. For each benchmark we give both Daisy and
Herbie a timeout of 2 minutes. To obtain our data we have run the script from
\autoref{fig:Daisy-Herbie-flow}.

\autoref{tbl:analysis_improvements1} and \autoref{tbl:analysis_improvements2} contain all benchmarks for which we can
compute an end-to-end result, i.e. at least one run of Daisy succeeds on both Herbies
input and output, and Herbie does not time out.

\autoref{tbl:rewriting_imprv1} and \autoref{tbl:rewriting_imprv2} contain all
benchmarks for which at least one of our rewritings applied in \autoref{sec:results}
succeds, i.e. either rewriting with only Daisy, only Herbie or with both does
not time out and does not raise an alarm.
In the table, \emph{TO} refers to either Herbie or Daisy not computing a
result within 2 minutes. \emph{SQRTNEG} and \emph{DIV0} refer to Daisy finding a
possible square root of a negative number, respectively division by zero.
\emph{POW} refers to Daisy encountering a power expression $a^b$ where $b$ is not
an integer value.
\emph{COND} and \emph{CBRT} refer to Daisy failing because the input contains
a conditional, respectively a cube root, as both are currently not implemented in Daisy.

In the following we list the benchmarks for which we did not obtain an end-to-end
result, sorted by the reason for the failure:
\begin{itemize}
\item Herbie timeout:\\
floudas1, hartman6, kepler1, kepler2,  triangle7, triangle8, triangle9,
triangle10, triangle11, triangle12, odometryX1, odometryY1, odometryX5,
pid3, pid5, leadLag3, leadLag5, trapezoid3, trapezoid5, train4\_out1,
train4\_state1, train4\_state2, train4\_state3, train4\_state4,
train4\_state5, train4\_state7, train4\_state8, train4\_state9

\item Daisy reports possible division by zero:\\
sine, pid1, NMSE problem 3.3.2, NMSE problem 3.3.3, NMSE problem 3.3.6,
NMSE p42 positive, NMSE p42 negative, NMSE problem 3.2.1 positive,
NMSE problem 3.2.1 negative, NMSE example 3.9, NMSE example 3.10,
NMSE problem 3.4.1, NMSE problem 3.4.2, NMSE problem 3.4.4, NMSE problem 3.4.5,
NMSE section 3.11

\item Herbie introduces cube-root which is unhandled by Daisy:\\
  arcLength, jetEngine, odometryX3, NMSE example 3.6, NMSE example 3.8

\item Daisy reports possible square root of a negative number:\\
triangle2, triangle3, triangle4, triangle5, triangle6

\item Use of non-integer power, unhandled by Daisy:\\
  NMSE problem 3.3.1

\item Internal exception by Daisy due to a conditional introduced by Herbie:\\
  NMSE problem 3.3.7

\end{itemize}

{\scriptsize
\begin{sidewaystable}
  \centering
\begin{tabular}{lccccccccccccc}
  \toprule
Name &\multicolumn{2}{c}{Herbie}  & \multicolumn{2}{c}{Daisy IA}  & \multicolumn{2}{c}{Daisy IA-subdiv} & \multicolumn{2}{c}{Daisy dReal}  & \multicolumn{2}{c}{Daisy Z3}  & \multicolumn{2}{c}{Daisy Best} & \shortstack{Error\\ improvement}\\
\midrule
 & src error &res error &src error &res error & src error & res error & src error &res error & src error & res error & src err & res err & res / src\\
 \midrule
logexp & 0.0289 & 0.0289 & 3.33e-12 & 3.33e-12 & 2.18e-13 & 2.18e-13 & 3.33e-12 & 3.33e-12 & 3.33e-12 & 3.33e-12 & 2.18e-13 & 2.18e-13 & 1.00 \\
sphere & 0.0105 & 0.0105 & 1.20e-14 & 1.20e-14 & 1.20e-14 & 1.20e-14 & 1.20e-14 & 1.20e-14 & 1.20e-14 & 1.20e-14 & 1.20e-14 & 1.20e-14 & 1.00 \\
floudas2 & 0 & 0 & 1.11e-15 & 1.11e-15 & 1.11e-15 & 1.11e-15 & 1.11e-15 & 1.11e-15 & 1.11e-15 & 1.11e-15 & 1.11e-15 & 1.11e-15 & 1.00 \\
floudas3 & 0.136 & 0.136 & 1.58e-14 & 1.58e-14 & 1.49e-14 & 1.49e-14 & 1.09e-14 & 1.09e-14 & 1.09e-14 & 1.09e-14 & 1.09e-14 & 1.09e-14 & 1.00 \\
hartman3 & 0.128 & 0.12 & 6.52e-10 & 6.52e-10 & 6.88e-14 & 6.88e-14 & 1.69e-13 & 1.69e-13 & 1.70e-13 & 1.70e-13 & 6.88e-14 & 6.88e-14 & 1.00 \\
kepler0 & 0.576 & 0.423 & 1.04e-13 & 6.54e-14 & 8.94e-14 & 5.42e-14 & 9.20e-14 & 6.54e-14 & 9.06e-14 & 6.54e-14 & 8.94e-14 & 5.42e-14 & 6.06e-01 \\
doppler1 & 0.434 & 0.434 & 4.19e-13 & 4.19e-13 & 2.38e-13 & 2.38e-13 & 4.19e-13 & 4.19e-13 & 4.19e-13 & 4.19e-13 & 2.38e-13 & 2.38e-13 & 1.00 \\
doppler2 & 0.424 & 0.0415 & 1.05e-12 & 4.25e-13 & 5.08e-13 & 3.43e-13 & 1.03e-12 & 3.63e-13 & 1.03e-12 & 3.63e-13 & 5.08e-13 & 3.43e-13 & 6.75e-01 \\
doppler3 & 0.423 & 0.0459 & 1.68e-13 & 1.06e-13 & 1.03e-13 & 9.55e-14 & 1.68e-13 & 9.85e-14 & 1.68e-13 & 9.85e-14 & 1.03e-13 & 9.55e-14 & 9.27e-01 \\
rigidBody1 & 0.00502 & 0.00577 & 3.22e-13 & 2.77e-13 & 3.22e-13 & 2.77e-13 & 3.22e-13 & 2.77e-13 & 3.22e-13 & 2.77e-13 & 3.22e-13 & 2.77e-13 & 8.60e-01 \\
rigidBody2 & 0.0658 & 0.0658 & 3.65e-11 & 5.75e-13 & 3.65e-11 & 5.75e-13 & 3.65e-11 & 5.75e-13 & 3.65e-11 & 5.75e-13 & 3.65e-11 & 5.75e-13 & 1.58e-02 \\
turbine1 & 0.429 & 0.511 & 9.49e-14 & 1.58e-13 & 4.24e-14 & 5.84e-14 & 8.87e-14 & 1.34e-13 & 8.87e-14 & 1.34e-13 & 4.24e-14 & 5.84e-14 & 1.38 \\
turbine2 & 0.554 & 0.66 & 1.39e-13 & 1.12e-10 & 4.58e-14 & 9.35e-13 & 1.23e-13 & 8.83e-11 & 1.23e-13 & 8.83e-11 & 4.58e-14 & 9.35e-13 & 2.04e+01 \\
turbine3 & 0.335 & 0.219 & 7.07e-14 & 3.26e-11 & 1.86e-14 & 6.96e-14 & 6.27e-14 & 3.26e-11 & 6.27e-14 & 3.26e-11 & 1.86e-14 & 6.96e-14 & 3.74 \\
verhulst & 0.336 & 0.385 & 4.74e-16 & 5.40e-16 & 3.59e-16 & 4.26e-16 & 4.18e-16 & 4.85e-16 & 4.18e-16 & 4.85e-16 & 3.59e-16 & 4.26e-16 & 1.19 \\
predatorPrey & 0.412 & 0.409 & 2.08e-16 & 2.39e-16 & 1.97e-16 & 2.28e-16 & 2.08e-16 & 2.39e-16 & 2.08e-16 & 2.39e-16 & 1.97e-16 & 2.28e-16 & 1.16 \\
carbonGas & 0.44 & 0.426 & 3.91e-08 & 1.66e-07 & 1.40e-08 & 2.54e-08 & 3.35e-08 & 1.60e-07 & 3.35e-08 & 1.60e-07 & 1.40e-08 & 2.54e-08 & 1.81 \\
sqroot & 0.008 & 0.00275 & 6.47e-16 & 4.54e-16 & 6.16e-16 & 3.58e-16 & 6.47e-16 & 4.23e-16 & 6.47e-16 & 4.22e-16 & 6.16e-16 & 3.58e-16 & 5.81e-01 \\
sineOrder3 & 0.00745 & 0.00745 & 1.45e-15 & 1.45e-15 & 1.01e-15 & 1.01e-15 & 1.23e-15 & 1.23e-15 & 1.23e-15 & 1.23e-15 & 1.01e-15 & 1.01e-15 & 1.00 \\
triangle & 2.55 & 0.455 & 6.53e-14 & 2.52e-14 & 5.58e-14 & 2.15e-14 & 6.05e-14 & 2.44e-14 & 6.04e-14 & 2.44e-14 & 5.58e-14 & 2.15e-14 & 3.85e-01 \\
triangle1 & 1.73 & 0.404 & SQRT & SQRT & SQRT & SQRT & 2.12e-11 & 1.11e-11 & 4.04e-10 & 4.88e-11 & 2.12e-11 & 1.11e-11 & 5.24e-01 \\
bspline3 & 0.116 & 0.0891 & 1.06e-16 & 1.06e-16 & 1.06e-16 & 1.06e-16 & 1.06e-16 & 1.06e-16 & 1.06e-16 & 1.06e-16 & 1.06e-16 & 1.06e-16 & 1.00 \\
\bottomrule
\end{tabular}
\caption{Roundoff error estimates computed by Herbie and all analysis run in Daisy. The ``Daisy Best'' columns are minima of Daisys src and res error across the different analysis, Error improvement is the error of Herbies result divided by the error of the input, numbers smaller than 0 mean that the worst-case roundoff error has been improved by Herbie - Part 1}
\label{tbl:analysis_improvements1}
\end{sidewaystable}
}

{\scriptsize
\begin{sidewaystable}
  \centering
\begin{tabular}{lccccccccccccc}
  \toprule
Name &\multicolumn{2}{c}{Herbie}  & \multicolumn{2}{c}{Daisy IA}  & \multicolumn{2}{c}{Daisy IA-subdiv} & \multicolumn{2}{c}{Daisy dReal}  & \multicolumn{2}{c}{Daisy Z3}  & \multicolumn{2}{c}{Daisy Best} & \shortstack{Error\\ improvement}\\
\midrule
 & src error &res error &src error &res error & src error & res error & src error &res error & src error & res error & src err & res err & res / src\\
 \midrule
pendulum1 & 0.0194 & 0.0194 & 4.61e-16 & 4.61e-16 & 4.61e-16 & 4.61e-16 & 4.61e-16 & 4.61e-16 & 4.61e-16 & 4.61e-16 & 4.61e-16 & 4.61e-16 & 1.00 \\
pendulum2 & 0.137 & 0.137 & 9.42e-16 & 9.38e-16 & 9.37e-16 & 9.33e-16 & 9.42e-16 & 9.38e-16 & 9.42e-16 & 9.38e-16 & 9.37e-16 & 9.33e-16 & 9.96e-01 \\
analysis1 & 0.0184 & 0.0184 & 1.67e-15 & 1.67e-15 & 9.41e-16 & 9.41e-16 & 1.30e-15 & 1.30e-15 & 1.67e-15 & 1.67e-15 & 9.41e-16 & 9.41e-16 & 1.00 \\
analysis2 & 0.0139 & 0.0139 & 6.08e-14 & 6.08e-14 & 2.07e-15 & 2.07e-15 & 3.11e-15 & 3.11e-15 & 6.08e-14 & 6.08e-14 & 2.07e-15 & 2.07e-15 & 1.00 \\
odometryY3 & 0.548 & 0.591 & 1.99e-14 & 1.91e-14 & 1.98e-14 & 1.89e-14 & 1.99e-14 & 1.91e-14 & 1.99e-14 & 1.91e-14 & 1.98e-14 & 1.89e-14 & 9.55e-01 \\
odometryY5 & 0.602 & 0.585 & 4.29e-14 & 3.88e-14 & 4.18e-14 & 3.70e-14 & 4.26e-14 & 3.76e-14 & 4.26e-14 & 3.76e-14 & 4.18e-14 & 3.70e-14 & 8.85e-01 \\
leadLag1 & 0.48 & 0 & 9.75e-12 & 3.20e-12 & 8.84e-12 & 3.20e-12 & 8.84e-12 & 3.20e-12 & 8.84e-12 & 3.20e-12 & 8.84e-12 & 3.20e-12 & 3.62e-01 \\
trapezoid1 & 5.49 & 0.365 & 6.58e-09 & 7.74e-11 & 6.58e-09 & 7.74e-11 & 6.58e-09 & 7.74e-11 & 6.58e-09 & 7.74e-11 & 6.58e-09 & 7.74e-11 & 1.18e-02 \\
intro-example & 0.022 & 0.022 & 1.14e-10 & 1.14e-10 & 2.84e-11 & 2.84e-11 & 1.14e-10 & 1.14e-10 & 1.14e-10 & 1.14e-10 & 2.84e-11 & 2.84e-11 & 1.00 \\
sec4-example & 0.404 & 0.305 & 2.49e-09 & 4.99e-16 & 1.43e-10 & 2.86e-16 & 2.49e-09 & 4.99e-16 & 2.49e-09 & 4.99e-16 & 1.43e-10 & 2.86e-16 & 2.00e-06 \\
test01\_sum3 & 0.361 & 0.144 & 4.00e-15 & 1.89e-15 & 3.55e-15 & 1.83e-15 & 3.55e-15 & 1.89e-15 & 3.55e-15 & 1.89e-15 & 3.55e-15 & 1.83e-15 & 5.15e-01 \\
test02\_sum8 & 0.416 & 0.349 & 7.99e-15 & 7.55e-15 & 7.99e-15 & 7.55e-15 & 7.99e-15 & 7.55e-15 & 7.99e-15 & 7.55e-15 & 7.99e-15 & 7.55e-15 & 9.45e-01 \\
test03\_nonlin2 & 0.0362 & 0.0415 & 4.75e-14 & 2.23e-11 & 1.28e-14 & 7.56e-13 & 4.67e-14 & 2.23e-11 & 4.67e-14 & 2.23e-11 & 1.28e-14 & 7.56e-13 & 5.91e+01 \\
test04\_dqmom9 & 0.628 & 0.499 & 2 & 4.21e+04 & 1.55 & 1.20e+04 & 2 & 4.21e+04 & 2 & 4.21e+04 & 1.55 & 1.20e+04 & 7.74e+03 \\
test05\_nonlin1 r4 & 0.776 & 0.343 & 3.89e-06 & 1.67e-16 & 6.32e-07 & 1.67e-16 & 3.89e-06 & 1.67e-16 & 3.89e-06 & 1.67e-16 & 6.32e-07 & 1.67e-16 & 2.64e-10 \\
test05\_nonlin1 test2 & 0.325 & 0.325 & 1.39e-16 & 1.39e-16 & 1.39e-16 & 1.39e-16 & 1.39e-16 & 1.39e-16 & 1.39e-16 & 1.39e-16 & 1.39e-16 & 1.39e-16 & 1.00 \\
test06\_sums4 sum1 & 0.00425 & 0.00425 & 1.33e-15 & 1.33e-15 & 1.33e-15 & 1.33e-15 & 1.33e-15 & 1.33e-15 & 1.33e-15 & 1.33e-15 & 1.33e-15 & 1.33e-15 & 1.00 \\
test06\_sums4 sum2 & 0.00363 & 0.00363 & 1.33e-15 & 1.33e-15 & 1.33e-15 & 1.33e-15 & 1.33e-15 & 1.33e-15 & 1.33e-15 & 1.33e-15 & 1.33e-15 & 1.33e-15 & 1.00 \\
NMSE example 3.1 & 50.1 & 0.435 & 0.041 & 6.49e-10 & 0.0102 & 3.09e-10 & 0.041 & 6.49e-10 & 0.041 & 6.49e-10 &   102e-02 & 3.09e-10 & 3.03e-08 \\
NMSE example 3.3 & 28.9 & 0.00575 & 1.33e-15 & 1.50e-25 & 7.77e-16 & 1.50e-25 & 1.11e-15 & 1.50e-25 & 1.33e-15 & 1.50e-25 & 7.77e-16 & 1.50e-25 & 1.93e-10 \\
NMSE example 3.4 & 61.9 & 0 & 2.22e+04 & 1.29e-26 & 2.22e+04 & 1.29e-26 & 2.22e+04 & 1.29e-26 & 2.22e+04 & 1.29e-26 & 2.22e+04 & 1.29e-26 & 5.81e-31 \\
NMSE problem 3.3.5 & 38.9 & 0.0404 & 8.88e-16 & 1.50e-25 & 7.77e-16 & 1.50e-25 & 6.67e-16 & 1.50e-25 & 8.88e-16 & 1.50e-25 & 6.67e-16 & 1.50e-25 & 2.25e-10 \\
NMSE example 3.7 & 60 & 0 & 2.22e-16 & 1.29e-26 & 2.22e-16 & 1.29e-26 & 2.22e-16 & 1.29e-26 & 2.22e-16 & 1.29e-26 & 2.22e-16 & 1.29e-26 & 5.81e-11 \\
NMSE problem 3.4.3 & 60 & 0 & 3.33e-16 & 5.17e-26 & 3.33e-16 & 5.17e-26 & 3.33e-16 & 5.17e-26 & 3.33e-16 & 5.17e-26 & 3.33e-16 & 5.17e-26 & 1.55e-10 \\
NMSE section 3.5 & 29.4 & 29.4 & 2.22e-16 & 2.22e-16 & 2.22e-16 & 2.22e-16 & 2.22e-16 & 2.22e-16 & 2.22e-16 & 2.22e-16 & 2.22e-16 & 2.22e-16 & 1.00 \\
\bottomrule
\end{tabular}
\caption{Roundoff error estimates computed by Herbie and all analysis run in Daisy. The ``Daisy Best'' columns are minima of Daisys src and res error across the different analysis, Error improvement is the error of Herbies result divided by the error of the input, numbers smaller than 0 mean that the worst-case roundoff error has been improved by Herbie - Part 2}
\label{tbl:analysis_improvements2}
\end{sidewaystable}
}

{\footnotesize
  \begin{table}
    \centering
  \begin{tabular}{cccccc}
    \toprule
Name & \shortstack{Baseline\\Daisy\\ without rewriting} & \multicolumn{3}{c}{Different Rewritings applied} \\
\midrule
 &  & Daisy & Herbie & Both & minimum \\
 \midrule
logexp & 3.33e-12 & 2.18e-13 & 3.33e-12 & 2.18e-13 & 2.18e-13 \\
sphere & 1.20e-14 & 1.17e-14 & 1.20e-14 & 1.17e-14 & 1.17e-14 \\
floudas1 & 3.01e-13 & 3.38e-13 & TO & TO & 3.38e-13 \\
floudas2 & 1.11e-15 & 1.11e-15 & 1.11e-15 & 1.11e-15 & 1.11e-15 \\
floudas3 & 1.09e-14 & 1.09e-14 & 1.09e-14 & 1.31e-14 & 1.09e-14 \\
hartman3 & 1.70e-13 & TO & 1.70e-13 & 6.92e-14 & 6.92e-14 \\
kepler0 & 9.06e-14 & 5.93e-14 & TO & TO & 5.93e-14 \\
kepler1 & 4.25e-13 & 2.70e-13 & TO & TO & 2.70e-13 \\
kepler2 & 2.19e-12 & 1.54e-12 & TO & TO & 1.54e-12 \\
doppler1 & 4.19e-13 & 1.62e-13 & 1.91e-13 & 1.47e-13 & 1.47e-13 \\
doppler2 & 1.03e-12 & 3.03e-13 & 3.63e-13 & 2.84e-13 & 2.84e-13 \\
doppler3 & 1.68e-13 & 7.97e-14 & 9.85e-14 & 7.17e-14 & 7.17e-14 \\
rigidBody1 & 3.22e-13 & 2.24e-13 & TO & TO & 2.24e-13 \\
rigidBody2 & 3.65e-11 & 2.94e-11 & 3.38e-11 & 3.35e-11 & 2.94e-11 \\
turbine1 & 8.87e-14 & 4.02e-14 & 9.02e-14 & 3.97e-14 & 3.97e-14 \\
turbine2 & 1.23e-13 & 4.51e-14 & 1.22e-13 & 4.50e-14 & 4.50e-14 \\
turbine3 & 6.27e-14 & 1.73e-14 & 3.27e-11 & 1.66e-14 & 1.66e-14 \\
verhulst & 4.18e-16 & 3.59e-16 & 4.85e-16 & 4.00e-16 & 3.59e-16 \\
predatorPrey & 2.08e-16 & 1.87e-16 & 5.92e-16 & 1.87e-16 & 1.87e-16 \\
carbonGas & 3.35e-08 & 1.20e-08 & 3.12e-08 & 1.11e-08 & 1.11e-08 \\
sine & 6.95e-16 & 4.86e-16 & DIV0 & DIV0 & 4.86e-16 \\
sqroot & 6.47e-16 & 2.35e-16 & 4.22e-16 & 2.35e-16 & 2.35e-16 \\
sineOrder3 & 1.23e-15 & 1.00e-15 & 1.23e-15 & 1.00e-15 & 1.00e-15 \\
triangle & 6.04e-14 & 5.58e-14 & 2.44e-14 & 2.08e-14 & 2.08e-14 \\
triangle1 & 4.04e-10 & SQRTNEG & 1.01e-06 & DIV0 & 1.01e-06 \\
triangle2 & SQRTNEG & SQRTNEG & 6.31e-10 & SQRTNEG & 6.31e-10 \\
bspline3 & 1.06e-16 & 1.06e-16 & 1.06e-16 & 1.06e-16 & 1.06e-16 \\
pendulum1 & 4.61e-16 & 4.61e-16 & 4.61e-16 & 4.60e-16 & 4.60e-16 \\
pendulum2 & 9.42e-16 & 9.36e-16 & 9.46e-16 & 9.37e-16 & 9.36e-16 \\
analysis1 & 1.67e-15 & 9.41e-16 & 1.67e-15 & 9.41e-16 & 9.41e-16 \\
analysis2 & 6.08e-14 & 2.07e-15 & 6.57e-14 & TO & 2.07e-15 \\
odometryX1 & 6.40e-15 & TO & 4.97e-15 & TO & 4.97e-15 \\
odometryY1 & 2.56e-15 & TO & 2.03e-15 & TO & 2.03e-15 \\
odometryX3 & 2.23e-14 & TO & 2.24e-14 & TO & 2.24e-14 \\
odometryY3 & 1.99e-14 & TO & 2.15e-14 & TO & 2.15e-14 \\
pid1 & 2.17e-15 & 2.17e-15 & FN/cbrt & FN/cbrt & 2.17e-15 \\
leadLag1 & 8.84e-12 & 8.84e-12 & 3.20e-12 & 3.20e-12 & 3.20e-12 \\
trapezoid1 & 6.58e-09 & TO & 5.14e-11 & 3.68e-11 & 3.68e-11 \\
trapezoid5 & 6.64e-09 & TO & 1.26e-10 & 7.15e-11 & 7.15e-11 \\
intro-example & 1.14e-10 & 2.84e-11 & DIV0 & DIV0 & 2.84e-11 \\
\bottomrule
  \end{tabular}
  \caption{Roundoff error estimates computed by Daisy for our evaluation of different rewriting approaches. The first column contains the baseline, Daisy refers to Daisy with Z3 and rewriting, Herbie refers to running Herbies optimizations only and ``Both'' means running first Herbie and then optimizing using Daisy with Z3 and rewriting. The column minimum gives the minimum roundoff error by any of the three optimizations - Part 1}
  \label{tbl:rewriting_imprv1}
\end{table}}

{\footnotesize
  \begin{table}
    \centering
  \begin{tabular}{cccccc}
    \toprule
Name & \shortstack{Baseline\\Daisy\\ without rewriting} & \multicolumn{3}{c}{Different Rewritings applied} \\
\midrule
 &  & Daisy & Herbie & Both & minimum \\
 \midrule
sec4-example & 2.49e-09 & 1.43e-10 & 2.49e-09 & 1.48e-10 & 1.43e-10 \\
test01\_sum3 & 3.55e-15 & 2.28e-15 & 1.89e-15 & 1.83e-15 & 1.83e-15 \\
test02\_sum8 & 7.99e-15 & 6.66e-15 & 7.55e-15 & 6.66e-15 & 6.66e-15 \\
test03\_nonlin2 & 4.67e-14 & 1.28e-14 & 4.67e-14 & 1.28e-14 & 1.28e-14 \\
test04\_dqmom9 & 2 & 1.55 & 4.21e+04 & 6.66e-01 & 6.66e-01 \\
test05\_nonlin1 r4 & 3.89e-06 & 6.32e-07 & 1.67e-16 & 1.39e-16 & 1.39e-16 \\
test05\_nonlin1 test2 & 1.39e-16 & 1.39e-16 & 2.78e-16 & 2.22e-16 & 1.39e-16 \\
test06\_sums4 sum1 & 1.33e-15 & 1.33e-15 & 1.33e-15 & 1.33e-15 & 1.33e-15 \\
test06\_sums4 sum2 & 1.33e-15 & 1.33e-15 & 1.33e-15 & 1.33e-15 & 1.33e-15 \\
NMSE example 3.1 & 0.041 & 1.02e-02 & 3.07e-12 & 2.30e-12 & 2.30e-12 \\
NMSE example 3.3 & 1.33e-15 & 7.77e-16 & 1.50e-25 & 1.39e-25 & 1.39e-25 \\
NMSE example 3.4 & 22200 & 2.22e+04 & 1.29e-26 & 9.69e-27 & 9.69e-27 \\
NMSE example 3.6 & 4.10e-12 & 1.02e-12 & DIV0 & DIV0 & 1.02e-12 \\
NMSE problem 3.3.1 & 8.19e-17 & 2.05e-17 & POW & POW & 2.05e-17 \\
NMSE problem 3.3.2 & DIV0 & DIV0 & 1.29e-26 & 1.29e-26 & 1.29e-26 \\
NMSE problem 3.3.3 & 1.64e-16 & 4.10e-17 & DIV0 & DIV0 & 4.10e-17 \\
NMSE problem 3.3.5 & 8.88e-16 & 7.77e-16 & 1.50e-25 & 1.39e-25 & 1.39e-25 \\
NMSE problem 3.3.6 & 8.19e-07 & 2.05e-07 & DIV0 & DIV0 & 2.05e-07 \\
NMSE problem 3.3.7 & 9.12e-16 & 9.66e-16 & COND & COND & 9.66e-16 \\
NMSE example 3.7 & 2.22e-16 & 2.22e-16 & 1.29e-26 & 1.29e-26 & 1.29e-26 \\
NMSE example 3.8 & 1.64e+14 & 6.83e+12 & FN/cbrt & FN/cbrt & 6.83e+12 \\
NMSE example 3.9 & DIV0 & DIV0 & 1.14e-26 & 1.14e-26 & 1.14e-26 \\
NMSE example 3.10 & DIV0 & DIV0 & 2.22e-16 & 1.11e-16 & 1.11e-16 \\
NMSE problem 3.4.3 & 3.33e-16 & 3.33e-16 & 5.17e-26 & 3.88e-26 & 3.88e-26 \\
NMSE problem 3.4.4 & DIV0 & DIV0 & 3.47e-16 & 2.68e-16 & 2.68e-16 \\
NMSE problem 3.4.5 & DIV0 & DIV0 & 1.11e-16 & 5.55e-17 & 5.55e-17 \\
NMSE section 3.5 & 2.22e-16 & 2.22e-16 & 3.55e-36 & 2.80e-36 & 2.80e-36 \\
NMSE section 3.11 & DIV0 & DIV0 & 6.46e+13 & 6.46e+13 & 6.46e+13 \\
train4\_out1 & 4.28e-10 & 3.19e-10 & 4.28e-10 & 3.19e-10 & 3.19e-10 \\
train4\_state1 & 6.66e-15 & 1.33e-15 & TO & TO & 1.33e-15 \\
train4\_state2 & 7.11e-15 & 8.90e-16 & TO & TO & 8.90e-16 \\
train4\_state3 & 6.66e-15 & 8.90e-16 & TO & TO & 8.90e-16 \\
train4\_state4 & 6.22e-15 & 1.33e-15 & TO & TO & 1.33e-15 \\
train4\_state5 & 1.23e-14 & 4.40e-15 & TO & TO & 4.40e-15 \\
train4\_state6 & 1.13e-14 & 4.22e-15 & TO & TO & 4.22e-15 \\
train4\_state7 & 1.04e-14 & 4.22e-15 & TO & TO & 4.22e-15 \\
train4\_state8 & 9.55e-15 & 4.22e-15 & TO & TO & 4.22e-15 \\
train4\_state9 & 8.66e-15 & 3.33e-15 & TO & TO & 3.33e-15 \\
\bottomrule
  \end{tabular}
  \caption{Roundoff error estimates computed by Daisy for our evaluation of different rewriting approaches. The first column contains the baseline, Daisy refers to Daisy with Z3 and rewriting, Herbie refers to running Herbies optimizations only and ``Both'' means running first Herbie and then optimizing using Daisy with Z3 and rewriting. The column minimum gives the minimum roundoff error by any of the three optimizations - Part 2}
  \label{tbl:rewriting_imprv2}
\end{table}}

\end{document}